\documentclass{PoS}

\title{Cosmogenic neutrinos and gamma-rays and the redshift evolution of UHECR sources}

\ShortTitle{Cosmogenic $\nu$ and $\gamma$ and UHECR source evol.}

\author{Roberto Aloisio\\
        Gran Sasso Science Institute (INFN), L'Aquila, Italy and\\
        INAF/Osservatorio Astrofisico di Arcetri, Firenze, Italy}

\author{Denise Boncioli\thanks{Now at DESY, Zeuthen, Germany}\\
        INFN/Laboratori Nazionali Gran Sasso, Assergi, Italy}

\author{\speaker{Armando di Matteo}\thanks{Now at ULB, Brussels, Belgium}\\
        INFN and Department of Physical and Chemical Sciences, Univ.~of L'Aquila, L'Aquila, Italy\\
        E-mail: \email{armando.dimatteo@aquila.infn.it}}

\author{Carmelo Evoli\\
        Gran Sasso Science Institute (INFN), L'Aquila, Italy}

\author{Aurelio F. Grillo\\
        INFN/Laboratori Nazionali Gran Sasso, Assergi, Italy}

\author{Sergio Petrera\\
        Gran Sasso Science Institute (INFN), L'Aquila, Italy and\\
        INFN and Department of Physical and Chemical Sciences, Univ.~of L'Aquila, L'Aquila, Italy}

\abstract{If ultra-high-energy cosmic rays (UHECRs) have extragalactic origins, as is widely assumed to be the case at least for the majority of cosmic rays with energies above a few EeV, secondary neutrinos and photons can be expected to be produced during the propagation of UHECRs through intergalactic space via interactions with cosmic background photons.  The fluxes of such secondary particles are strongly dependent on the redshift evolution of the emissivity (number density times luminosity) of UHECR sources.  We show how cosmic rays, neutrinos, and gamma rays can potentially provide complementary information about UHECR source evolution.}

\FullConference{Neutrino Oscillation Workshop\\
		4 - 11 September, 2016\\
		Otranto (Lecce, Italy)}

\usepackage{caption,lineno}
\begin{document}

\section{Ultra-high-energy cosmic rays}
Particles of extraterrestrial origins with energies higher than $10^{18}$~eV are known as ultra-high-energy cosmic rays (UHECRs).
All or almost all these particles are protons and possibly other atomic nuclei, with stringent upper limits on the fluxes of non-hadronic particles such as photons or neutrinos. The origin of UHECRs is still unknown, but there is a wide consensus that most of the highest-energy cosmic rays originate from extragalactic sources.

The energy spectrum of UHECRs 
is in first approximation a power-law, $\mathrm{d}N/\mathrm{d}E \approx E^{-3}$, but there is an ``ankle'' feature around 5~EeV where the spectral index changes from 3.3 to~2.6 and a cutoff around 40~EeV above which the flux is steeply suppressed. Modern experiments agree about the position of the ankle within their uncertainties, but there are discrepancies in the measured positions of the cutoff.

\subsection{Propagation through intergalactic space}
Particles of extragalactic origin undergo various processes during their travel through intergalactic space which substantially alter their properties.  All particles are affected by an adiabatic energy loss due to the expansion of the universe.  Protons and nuclei can interact with photons of the cosmic microwave background (CMB) and the infrared/visible/ultraviolet extragalactic background light (EBL), such as pair photoproduction, photodisintegration, and pion photoproduction, losing energy and producing secondary particles (as described below). The trajectories of charged particles can be deflected by intergalactic and galactic magnetic fields.

\subsection{The secondary particles produced}
Pion photoproduction, e.g.~$\mathrm{p}+\gamma \to \mathrm{p}+\pi^0$, mainly affects nucleons with energies above 40~EeV with CMB photons, or, with a much lower interaction rate, nucleons above 4~EeV with EBL photons. Neutral pions then decay into photons, each with $O(10\%)$ of the initial nucleon energy, which undergo electron--positron pair production with CMB photons initiating electromagnetic cascades of TeV photons, whereas charged pions decay into neutrinos, each with $O(5\%)$ of the initial nucleon energy, which can reach Earth unaffected by the propagation except for the redshift energy loss and flavour oscillations.

Photodisintegration of nuclei, e.g.~${^{A}Z} + \gamma \to {^{A-1}Z} + \mathrm{n}$, mainly affects nuclei with energies per nucleon above 2~EeV with CMB photons and  nuclei with energies per nucleon above 0.2~EeV with EBL photons. It has important effects on energy spectrum and mass composition of UHE nuclei, but few direct multi-messenger implications.

Pair photoproduction, i.e.~$N + \gamma \to N + \mathrm{e}^+ + \mathrm{e}^-$, mainly affects nuclei (including protons) with energies per nucleon above 0.2~EeV with CMB photons, resulting in electrons and positrons each with $O(0.05\%)$ of the initial nucleus energy. These undergo inverse Compton scattering with background photons or synchrotron ratiation with intergalactic magnetic fields, initiating electromagnetic cascades of TeV photons. The shape of the energy spectrum of cascade photons at Earth is largely independent on the energy and charge of the initial electron or photon, and only depends on the redshift of its production point.

\subsection{Open questions}
Unanswered questions about UHECRs include the nature and distribution of their sources, the origin of the ankle (e.g.~the pair-production dip or a transition between two components) and the cutoff (e.g.~propagation effects and/or a maximum acceleration magnetic rigidity) in their energy spectrum, and their mass composition at the highest energies.

\section{Multi-messenger studies}
Due to propagation effects, no protons with energies above 1~EeV can originate from~$z>1$, 
and this limit is even more stringent for other nuclei.  On the other hand, secondary neutrinos from their propagation can reach Earth even from high redshifts, carrying information about remote UHECR sources.  Also, charged cosmic rays are deflected by magnetic fields, possibly by several tens of degrees, whereas neutral particles arrive from their original direction. (In principle, neutrinos carry more information than cascade gamma rays, but they are harder to detect.)
\subsection{Experimental limits on EeV fluxes and measured GeV--PeV fluxes}
\begin{figure}[h!]
  \centering
  \includegraphics[width=0.9\textwidth]{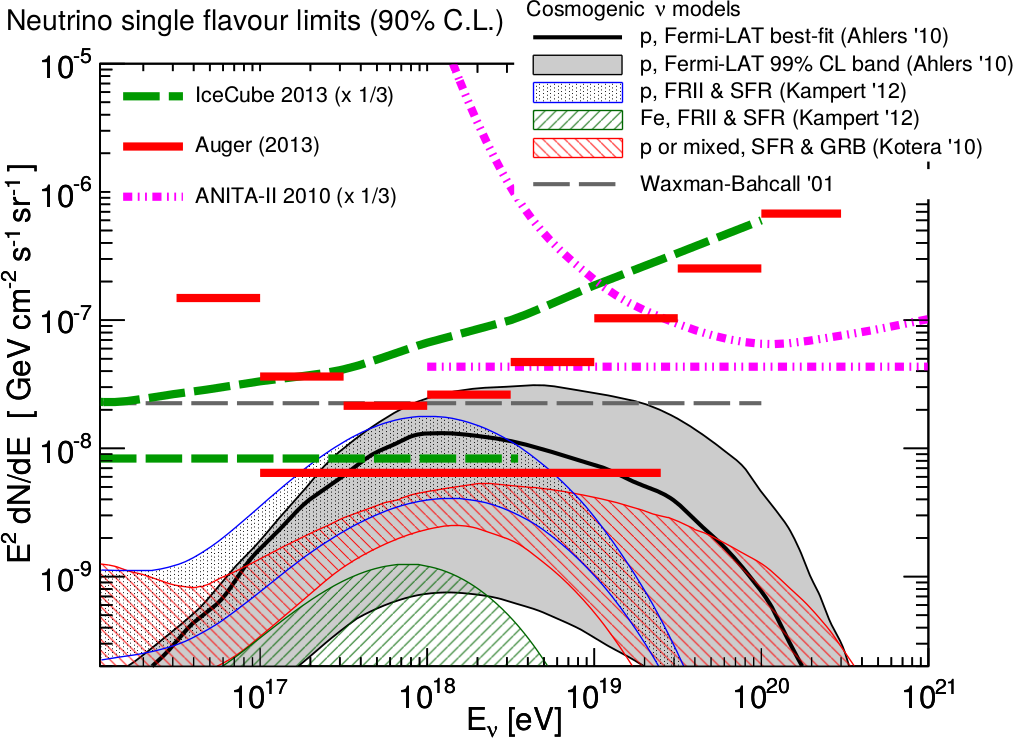} \includegraphics[width=0.9\textwidth]{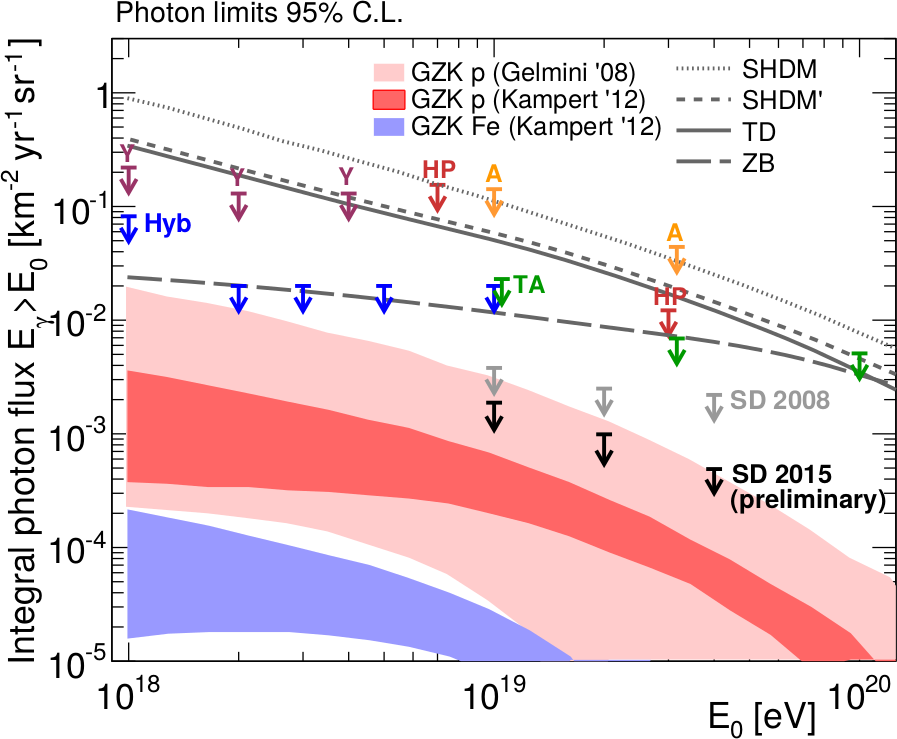}
  \caption{Upper limits on EeV neutrino~(top) and $\gamma$-ray~(bottom) fluxes by Auger~\cite{neutral}\label{fig:EeVlimits}}
\end{figure}

Upper limits on EeV neutrino and gamma-ray fluxes from Pierre Auger Observatory data are shown and compared with predictions from various models in Figure~\ref{fig:EeVlimits}.

\begin{figure}[h!]
  \centering
  \includegraphics[width=0.75\textwidth]{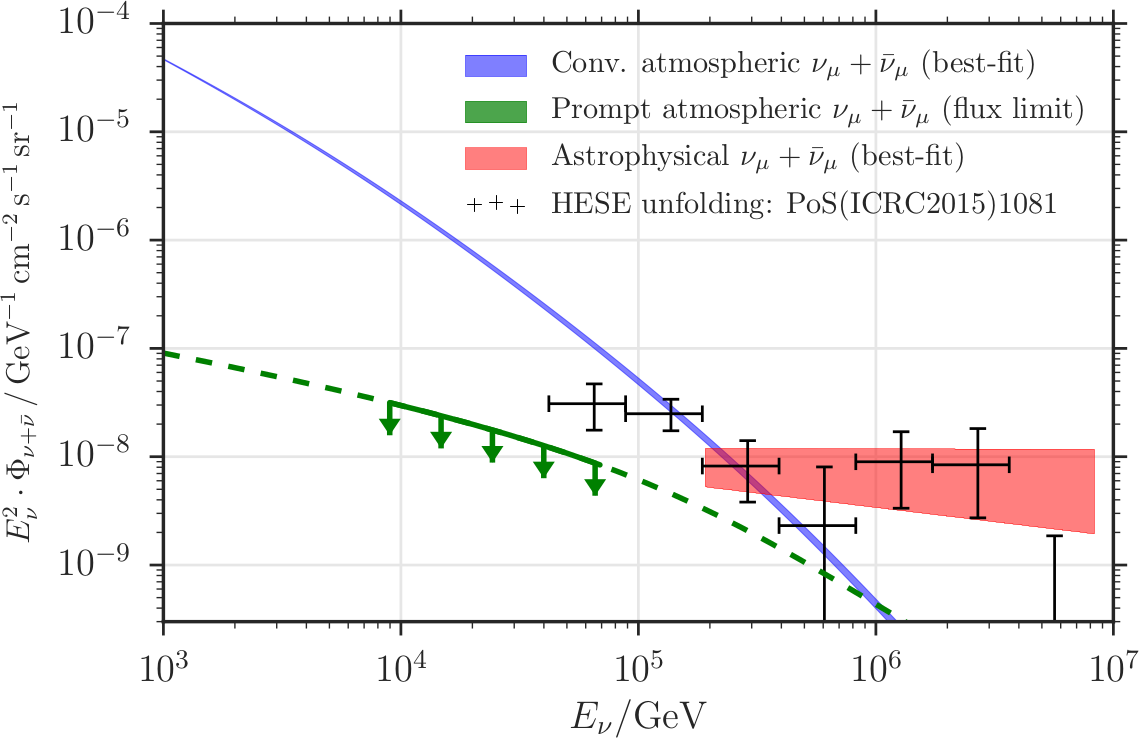} \includegraphics[width=0.75\textwidth]{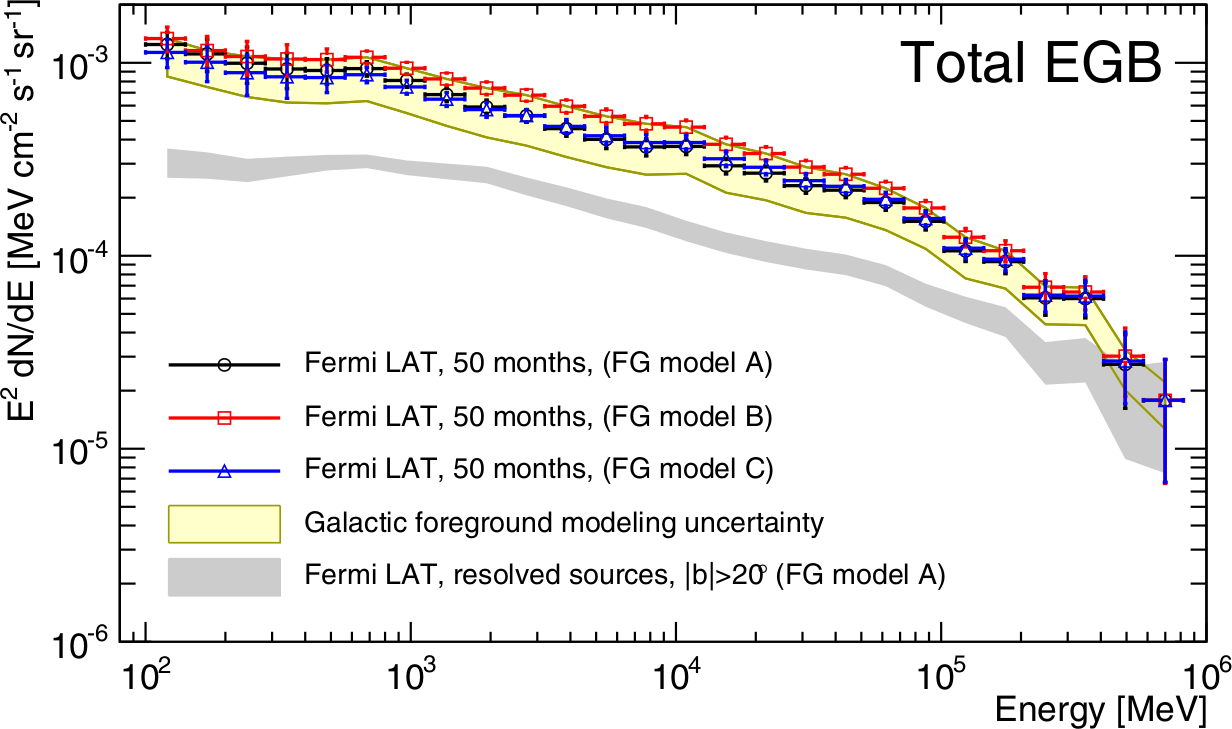}
  \caption{Measured PeV neutrino fluxes by IceCube~\cite{IceCube} (top) and GeV $\gamma$-ray fluxes by Fermi-LAT~\cite{FermiLAT} (bottom)\label{fig:TeVlimits}}
\end{figure}

The astrophysical neutrino flux at TeV--PeV energies measured by IceCube and the measured extragalactic gamma-ray background at GeV--TeV energies measured by Fermi-LAT are shown in Figure~\ref{fig:TeVlimits}.
\subsection{Simulated expected neutrino  and gamma-ray fluxes in various scenarios}
\begin{figure}
  \centering
  \includegraphics[width=0.45\textwidth]{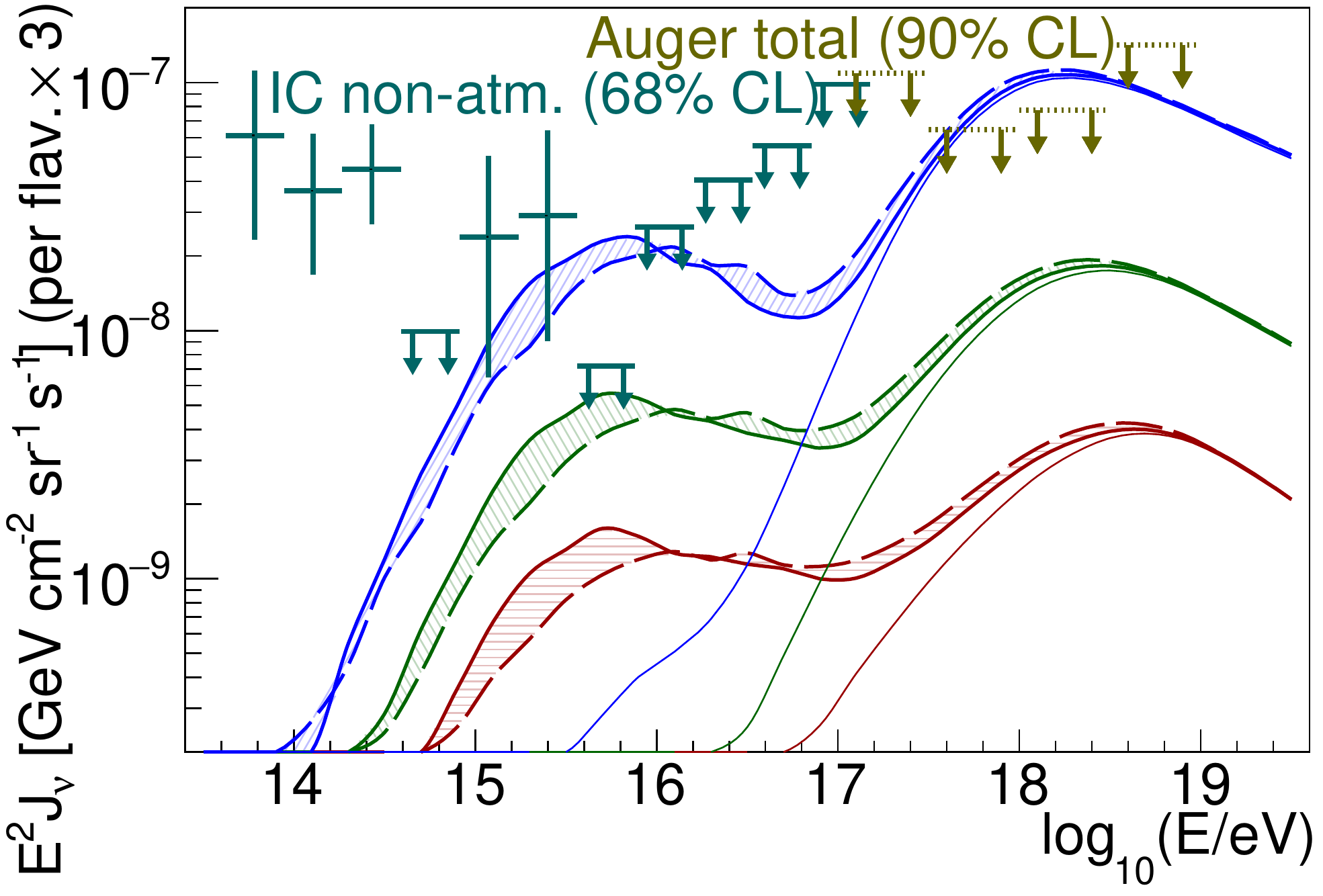}
  \includegraphics[width=0.45\textwidth]{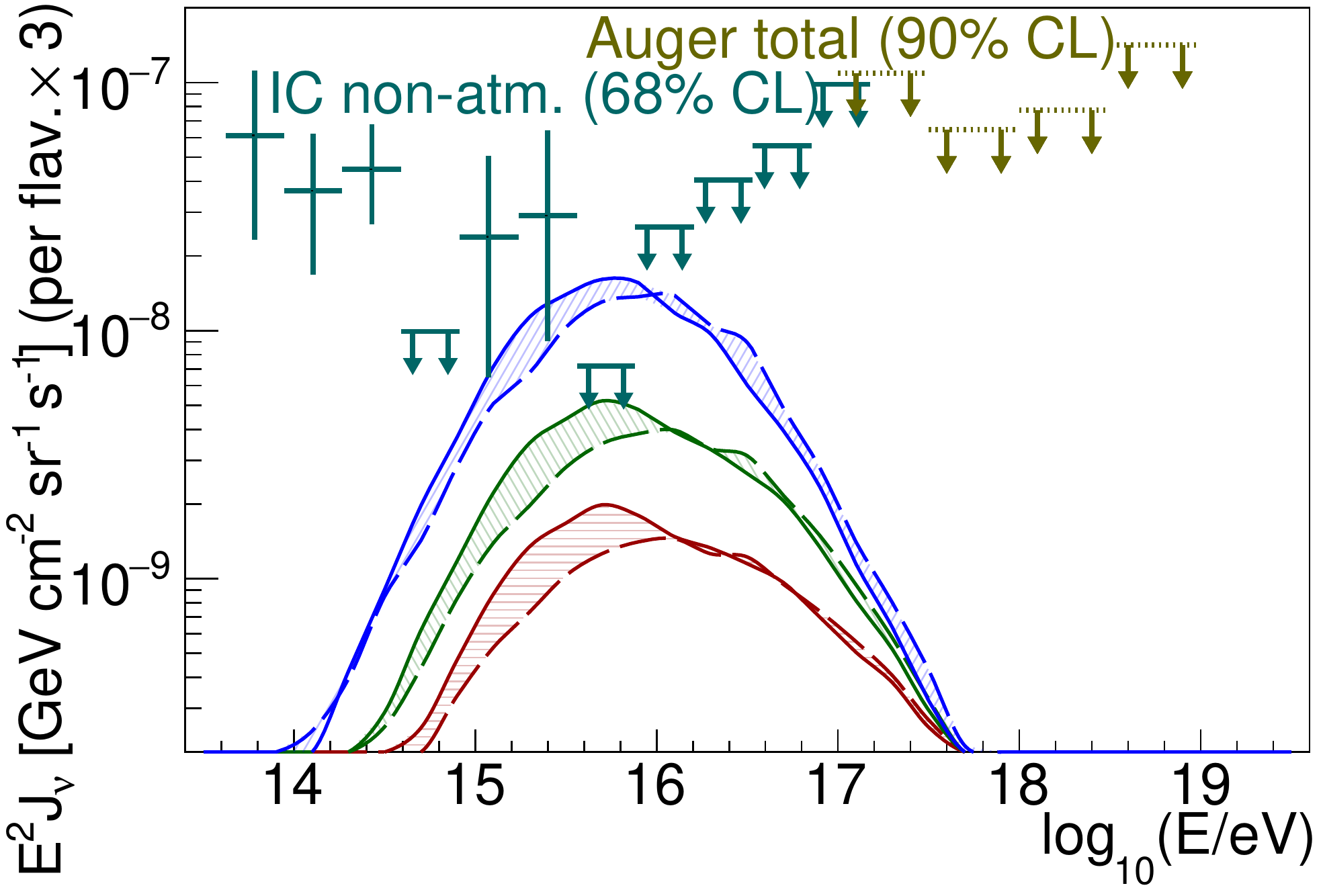}
  \caption{\label{fig:neutrinos}Expected cosmogenic neutrino flux in a proton-only UHECR scenario (right) and in a mixed-composition scenario (center), for various evolutions of source emissivity (from bottom to top: constant, SFR, and AGN), adapted from ref.~\cite{neutrinos}}
\end{figure}

In ref.~\cite{neutrinos}, we used \textit{SimProp}~v2r2 \cite{SimPropv2r2} to compute expected cosmogenic neutrino fluxes in various UHECR scenarios. The results are shown in the first two panels of Figure~\ref{fig:neutrinos} and show that scenarios where UHECRs are protons accelerated in sources whose emissivity increases with redshift are disfavored by the non-observation of neutrinos with energies above 10~PeV.

The upcoming \textit{SimProp} release \textit{SimProp}~v2r4 \cite{SimPropv2r4} will also output the secondary electrons and positrons from pair production, so that cascades can be computed with external tools such as ELMAG \cite{ELMAG}.  Similar studies by other authors \cite{Taylor,Kalashev} show that the resulting limits on UHECR source evolutions are qualitatively similar to those from neutrinos, but somewhat more stringent.

\section{Conclusions}
Cosmogenic neutrino fluxes depend on both the mass composition and the source evolution of UHECRs. EeV neutrinos can only be produced if there are protons among highest-energy CRs, and neutrino fluxes at all energies are strongly dependent of the emissivity of remote sources. 
We can already rule out models with source emissivity too strongly increasing with redshift (decreasing with time).
Similar considerations apply to gamma-ray fluxes; the interpretation is more complicated, but the limits that can be put on source emissivity are more stringent.

\providecommand{\href}[2]{#2}\begingroup\raggedright\endgroup
\clearpage

\end{document}